\documentclass[aps,preprint,nofootinbib]{revtex4}
\usepackage{graphicx}
\usepackage{amsmath}
\usepackage{amsfonts}
\usepackage{amssymb}

\begin{document}

\noindent{CITUSC/02-017 \ \hfill\ \ hep-th/0205194}\bigskip

\title{{\Large A Mysterious Zero in AdS$_{5}\times$S$^{5}$ Supergravity }}\bigskip

\author{\textbf{{\large {Itzhak Bars}}}{\footnote{
Research partially supported by the US Department of Energy under
grant number DE-FG03-84ER40168.}}}

\address{\large{Department of Physics and Astronomy\\
 University of Southern California,
Los Angeles, CA 90089-0484}} 

\begin{abstract}
It is shown that all the states in AdS$_{5}\times$S$^{5}$
supergravity have zero eigenvalue for all Casimirs of its symmetry
group SU$( 2,2|4)$ . To compute this zero in supergravity we
refine the oscillator methods for studying the lowest weight
unitary representations of SU$(N,M|R,S)$. We solve the reduction
problem when one multiplies an arbitrary number of super
doubletons. This enters in the computation of the Casimir
eigenvalues of the lowest weight representations. We apply the results to SU$%
(2,2|4)$ that classifies the Kaluza-Klein towers of ten
dimensional type IIB supergravity compactified on
AdS$_{5}\times$S$^{5}.$ We show that the vanishing of the
SU$(2,2|4)$ Casimir eigenvalues for all the states is indeed a
group theoretical fact in AdS$_{5}\times$S$^{5}$ supergravity. By
the AdS-CFT correspondence, it is also a fact for gauge invariant
states of super Yang-Mills theory with four supersymmetries in
four dimensions. This non-trivial and mysterious zero is very
interesting because it is predicted as a straightforward
consequence of the fundamental local Sp(2) symmetry in 2T-physics.
Via the 2T-physics explanation of this zero we find a global
indication that these special supergravity and super Yang-Mills
theories hide a twelve dimensional structure with (10,2)
signature.
\end{abstract}

\maketitle

\section{Introduction}

In recent sudies it has been shown that the non-linearly realized
hidden superconformal symmetries OSp$\left( N|4\right) ,$
SU$\left( 2,2|N\right) ,$ F$_{4},$ OSp$\left( 8^{\ast }|N\right) $
of ordinary superparticle actions in $d=3,4,5,6$ respectively,
become linearly realized and evident symmetries in the 2T-physics
formulation of superparticles \cite{super2t}\cite{survey2T} . The
2T-physics method has been extended to 12-dimensional super phase
space $\left( X^{M},P^{M},\Theta \right) $ with (10+2 signature,
and 32 dimensional Weyl spinor $\Theta $) to describe the
collection of all the Kaluza-Klein towers in AdS$_{5}\times
$S$^{5}$ supergravity, as the quantum states of a single SU$\left(
2,2|4\right) $ super particle on AdS$_{5}\times $S$^{5}$
\cite{survey2T}\cite{AdS5S5}$.$ This has predicted an interesting
property of type IIB supergravity that had not been noticed
before. Namely, the Casimir eigenvalues vanish for all the
SU$\left( 2,2|4\right) $ Casimir operators for all the states of
supergravity. The mechanism for this outcome in \cite{AdS5S5} is
the fundamental Sp$\left( 2\right) $ gauge symmetry and kappa
supersymmetry in 2T-physics. Namely, the gauge singlet condition
(vanishing of 12-dimensional constraints) on physical states
predicts the vanishing of the SU$\left( 2,2|4\right) $ Casimir
eigenvalues. From the point of view of supergravity this is rather
mysterious and requires verification with other methods. This
motivated the present study of representation theory of the
supergroup SU$\left( N,M|R,S\right) $ to compute the corresponding
Casimir eigenvalues with \textquotedblleft oscillator
methods\textquotedblright\ that were previously applied in the
studies of compactification of type IIB supergravity in 10
dimensions. In this paper all the Casimir eigenvalues are computed
and verified to vanish.

The following method for determining lowest weight unitary
representations of noncompact supergroups was developed in
\cite{barsgunaydin}. Here we refine the method to get a clearer
description of the lowest states in unitary representations of
SU$\left( N,M|R,S\right) $, and to compute the Casimir eigenvalues
for such representations, which were not properly computed in
previous literature. The correct computation of the Casimir
eigenvalues is essential in our application, and is needed to
prove that all the AdS$_{5}\times$S$^{5}$ supergravity states have
zero SU$\left( 2,2|4\right) $ Casimir eigenvalues.

Our aim is to study the case of SU$\left( 2,2|4\right) $ which is the
relevant supergroup. In the case of SU$\left( 2,2|4\right) $ the SU$\left(
2,2\right) $ subgroup is interpreted as the conformal group SO$\left(
4,2\right) $ which is the symmetry of AdS$_{5}$ space. Likewise the SU$%
\left( 4\right) =$SO$\left( 6\right) $ subgroup is interpreted as the
symmetry of S$^{5}$ space. The SU$\left( 2\right) \times$SU$\left( 2\right)
\times$SU$\left( 4\right) $ quantum numbers of the lowest states identify
the fields that enter in a field theory. We keep this physical application
in mind as we discuss first the more general setting of SU$\left(
N,M|R,S\right) ,$ and later specialize to SU$\left( 2,2|4\right) .$

Using the \textquotedblleft color\textquotedblright\ terminology of \cite%
{barsgunaydin}, the number of \textquotedblleft colors\textquotedblright\ in
this setup is interpreted as a device for solving the reduction problem when
one takes direct products of the fundamental unitary representation, which
is called super-doubleton for SU$(N,M|R,S)$ for one color. Thus the
reduction of the direct product of two doubletons is achieved by taking two
colors. For the direct product of three doubletons we take three colors, and
so on. It is natural to define a color group SU$\left( C\right) $ that acts
on $C$ copies of the super-doubleton. This trick of \textquotedblleft
color\textquotedblright\ was first used for compact supergroups to derive
the results in \cite{BB} and later applied to noncompact groups \cite{GS}
and supergroups \cite{barsgunaydin}.

We show that the \textquotedblleft color\textquotedblright\ group of the
oscillators cleanly classifies all the lowest states of SU$(N,M|R,S)$ and
determines the Casimir eigenvalues. The refinement in the methods of \cite%
{barsgunaydin} comes from improved use of the properties of the
\textquotedblleft color\textquotedblright\ group, which in turn permit the
new computation of the Casimir operator for these representations.

After classifying all the SU$(N,M|R,S)$ lowest states that appear in the
reduction according to the color group SU$\left( C\right) $, we focus on the
smaller set of color singlet representations that are of interest in our
physical application. We present a new and concise characterization of such
color singlet super multiplets in terms of a single lowest state which
identifies the supermultiplet of fields in field theory. In the case of SU$
\left( 2,2|4\right) $ we identify a unique color singlet multiplet for each
value of the color $C.$ This unique lowest state identifies all the fields
in a Kaluza-Klein tower that occurs in the compactification of 10D type IIB
supergravity on AdS$_{5}\times $S$^{5}.$ Taking all the values of color $%
C=2,3,\cdots ,\infty $ we find the supergravity set of Kaluza-Klein towers
which coincide with the quantum states of our superparticle in 12D in
2T-physics \cite{AdS5S5}. In the 2T-physics formulation of the problem the
vanishing of the SU$\left( 2,2|4\right) $ Casimir eigenvalues was a direct
consequence of Sp$\left( 2\right) $ and kappa gauge symmetries. In this
paper we verify this prediction through representation theory of SU$\left(
2,2|4\right) $ in the context of supergravity.

\section{Oscillator construction and the Casimir}

Consider the annihilation operators ($a_{n}^{\alpha},b_{\alpha m}$ = bosons,
$\psi_{r}^{\alpha},\chi_{\alpha s}$ = fermions), and the corresponding
creation operators (denoted by placing a bar on the symbol). The lower/upper
SU$\left( C\right) $ labels $\alpha=1,\cdots,C$ denote copies (named
\textquotedblleft color\textquotedblright\ in \cite{barsgunaydin}) and the
other indices belong to the fundamental representation of SU$\left( N\right)
\times$SU$\left( M\right) \times$SU$\left( R\right) \times$SU$\left(
S\right) $ as follows
\begin{align}
n,n^{\prime} & =1,2,\ldots,N;\;\;m,m^{\prime}=1,2,\ldots,M; \\
r,r^{\prime} & =1,2,\ldots,R;\;\;\;s,s^{\prime}=1,2,\ldots,S.
\end{align}
Lower indices are unprimed and upper indices are primed. They indicate the
fundamental representation (lower) and the complex conjugate of the
fundamental representation (upper) respectively. The generators of the
supergroups are constructed from the harmonic oscillators as follows (the
color index $\alpha$ is summed over, so the SU$\left( N,M|R,S\right) $
generators are color singlets)
\begin{align}
& \left(
\begin{array}{c}
a_{n} \\
\bar{b}^{m^{\prime}} \\
\psi_{r} \\
\bar{\chi}^{s^{\prime}}%
\end{array}
\right) ^{\alpha}\left(
\begin{array}{cccc}
\bar{a}^{n^{\prime}} & -b_{m} & \bar{\psi}^{r^{\prime}} & \chi_{s}%
\end{array}
\right) _{\alpha} \\
& =\left(
\begin{array}{cccc}
a_{n}\cdot\bar{a}^{n^{\prime}} & -a_{n}\cdot b_{m} & a_{n}\cdot\bar{\psi }
^{r^{\prime}} & a_{n}\cdot\chi_{s} \\
\bar{b}^{m^{\prime}}\cdot\bar{a}^{n^{\prime}} & -\bar{b}^{m^{\prime}}\cdot
b_{m} & \bar{b}^{m^{\prime}}\cdot\bar{\psi}^{r^{\prime}} & \bar{b}
^{m^{\prime }}\cdot\chi_{s} \\
\psi_{r}\cdot\bar{a}^{n^{\prime}} & -\psi_{r}\cdot b_{m} & \psi_{r}\cdot
\bar{\psi}^{r^{\prime}} & \psi_{r}\cdot\chi_{s} \\
\bar{\chi}^{s^{\prime}}\cdot\bar{a}^{n^{\prime}} & -\bar{\chi}
^{s^{\prime}}\cdot b_{m} & \bar{\chi}^{s^{\prime}}\cdot\bar{\psi}
^{r^{\prime}} & \bar {\chi}^{s^{\prime}}\cdot\chi_{s}%
\end{array}
\right) \\
& =J+\mathbf{1\,}\frac{\left( \Delta+\left( N-R\right) C\right) }{N+M-R-S}
\label{J}
\end{align}
The column is the fundamental of the supergroup, and the row is the
hermitian conjugate of the fundamental times the invariant metric, and the
matrix represents the generators in the adjoint. The dot represents
summation over the \textquotedblleft color\textquotedblright\ index $\alpha$%
. We are distinguishing between upper and lower indices. On the diagonal
blocks we separate the traceless and trace parts, and get the SU$\left(
N\right) ,$ SU$\left( M\right) ,$ SU$\left( R\right) ,$ SU$\left( S\right) $
subgroups and four U$\left( 1\right) $ subgroups generated by the number
operators $N_{a}=tr\left( \bar{a}\cdot a\right) ,$ $N_{b}=tr\left( \bar
{b}%
\cdot b\right) ,$ $N_{\psi}=tr\left( \bar{\psi}\cdot\psi\right) ,$ $%
N_{\chi}=tr\left( \bar{\chi}\cdot\chi\right) $. The supertraceless part of
the matrix denoted as $J$ gives the the SU$(N,M|R,S)$ generators. The
coefficient of the matrix $\mathbf{1}$ is the supertrace of the matrix,
while $\Delta$ is given in terms of the number operators by
\begin{equation}
\Delta=N_{a}-N_{b}+N_{\psi}-N_{\chi}.
\end{equation}
A SU$\left( C\right) \times$U$\left( 1\right) $ \textquotedblleft
color\textquotedblright\ group commutes with all the generators
$J$. The U$\left( 1\right) $ is generated by $\Delta$ and the
color SU$\left( C\right) $ is generated by the traceless matrix
$G_{\alpha}^{\,\,\beta}$
\begin{equation}
G_{\alpha}^{\,\,\beta}=\bar{a}_{\alpha}^{n}a_{n}^{\,\beta}-b_{\alpha m}\bar {%
b}^{m\beta}+\bar{\psi}_{\alpha}^{r}\psi_{r}^{\,\,\beta}+\chi_{\alpha s}\bar{
\chi}^{s\beta}-\delta_{\alpha}^{\,\,\beta}\frac{1}{C}\left(
\Delta-CM+CS\right) .
\end{equation}
The traceless $G_{\alpha}^{\,\,\beta}$ has the commutation rule $\left[
G_{\alpha}^{\,\,\beta},\Xi^{\gamma}\right] =-\delta_{\alpha}^{\gamma}\,\Xi^{
\alpha}$ (fundamental of \textquotedblleft color\textquotedblright) for
every component $\Xi^{\alpha}$ in the column matrix given above.

We can compute the quadratic Casimir operator of SU$\left(
N,M|R,S\right) $, which is given by the supertrace of the square
of $J$, as
\begin{equation}
C_{2}^{\left( N,M|R,S\right) }=\frac{1}{2}Str\left( JJ\right) .
\end{equation}
Higher Casimir operators are simply the supertrace of higher powers of $J.$
After some algebra one can verify our result that the the quadratic Casimir
operator is a function of only the Casimir operators of SU$\left( C\right)
\times$U$\left( 1\right) $
\begin{equation}
C_{2}^{\left( N,M|R,S\right) }=C_{2}^{SU\left( C\right) }+\frac{\left(
N+M-R-S\right) -C}{2\left( N+M-R-S\right) C}\left( \Delta+\left( N-R\right)
C\right) \left( \Delta-\left( M-S\right) C\right)  \label{C2}
\end{equation}
where $C_{2}^{SU\left( C\right) }$ is the quadratic Casimir for SU$\left(
C\right) $ given by%
\begin{equation}
C_{2}^{SU\left( C\right) }=\frac{1}{2}G_{\alpha}^{\,\,\beta}G_{\beta
}^{\,\,\beta}.
\end{equation}

The above expression for $C_{2}^{\left( N,M|R,S\right) }$ is
incorrect when $N+M=R+S$ because in that case the expression in
Eq.(\ref{J}) is not valid due to the fact that one must remove
from $J$ one generator proportional to the identity matrix which
is supertraceless. The relevant generator is identified as the
U$(1)$ in the decomposition
SU$(N,M|R,S)\rightarrow$SU$(N|R)\times$SU$(M|S)\times$U$(1)$,
which is proportional to
$\delta=C+(N_a+N_\psi)/(N-R)-(N_b+N_\chi)/(M-S)$. When $N+M=R+S$
it becomes $C+\Delta/(N-R)$ which commutes with all other
generators. Its contribution to the Str$(J^2)/2$ must be
subtracted to obtain the correct Casimir when $N+M=R+S$. This
amounts to subtracting the quantity
$(\delta)^2(N-R)(M-S)/2(N+M-R-S)$ from
$C_{2}^{\left(N,M|R,S\right) }$ and then taking the limit
$N+M=R+S$. When this is taken into account in the computation of
the quadratic Casimir, the result is
\begin{equation}
\left. C_{2}^{\left( N,M|R,S\right) }\right\vert _{N+M=R+S}=C_{2}^{SU\left(
C\right) }+\frac{1}{2C}\left( \Delta+\left( N-R\right) C\right) ^{2}+\frac{C
}{2}\left( \Delta+\left( N-R\right) C\right) .  \label{C2equal}
\end{equation}
Note that the pole at $N+M=R+S$ has cancelled in this expression.
Both expressions in Eqs.(\ref{C2},\ref{C2equal}) for the Casimir
operators in the oscillator approach for supergroups are new. We
will use them in our discussion.

In particular, for SU$\left( 2,2|4\right) $ color singlet states, which are
important in our setup below, Eq.(\ref{C2equal}) reduces to ($N=M=2,$ $R=4,$
$S=0,$ $C_{2}^{SU\left( C\right) }=0$)
\begin{equation}
C_{2}^{\left( 2,2|4\right) }=\frac{1}{2C}\left( \Delta-2C+C^{2}\right)
\left( \Delta-2C\right) .  \label{colorless}
\end{equation}
In the rest of the paper we will show that for all the states in
the Kaluza-Klein towers of supergravity on AdS$_{5}\times$S$^{5}$
we must have $\Delta=2C,$ for $C=2,3,\cdots\infty,$ and therefore
$C_{2}^{\left( 2,2|4\right) }=0$ in AdS$_{5}\times$S$^{5}$
supergravity.

One may check our general expression for the Casimir operator in
various limits of the numbers $M,N,R,S$ for which one can find
formulas in the literature. For example for SU$\left( N\right) $
with one color $C=1$ we know we will obtain only the one row Young
tableaux because of the symmetry imposed by the products of
bosonic creation operators $\bar{a}^{n_{1}^{\prime
}}\cdots\bar{a}^{n_{k}^{\prime}}|0>$. Our formula in Eq.(\ref{C2})
reduces then to the Casimir for SU$\left( N\right) $ when we take
the limit $M=R=S=0$ and use $C_{2}^{SU\left( 1\right) }=0$
\begin{equation}
\left. C_{2}^{SU\left( N\right) }\right\vert _{one-row}=\frac{N-1}{2N}
\left( \Delta+N\right) \Delta
\end{equation}
where $\Delta=N_{a}$ is the number of boxes. Indeed this is the correct
Casimir eigenvalue for the one row Young tableaux representations. In
particular for SU$\left( 2\right) $ we have $N=2,$ and defining $j=\Delta/2$
gives $C_{2}^{SU\left( 2\right) }=j\left( j+1\right) $ which is the well
known formula.

Another simple example is the case $N=M=S=0$ which leaves SU$\left( R\right)
$ constructed with fermionic oscillators $\psi_{r}$. In the case of one
color $C=1$ we can obtain only the one column antisymmetric representations
of SU$\left( R\right) $ because of the antisymmetry imposed by fermionic
creation operators $\bar{\psi}^{r_{1}^{\prime}}\cdots\bar{\psi}%
^{r_{k}^{\prime}}|0>$. In this case our formula reduces to (the minus sign
is because of the supertrace)%
\begin{equation}
-\left. C_{2}^{SU\left( R\right) }\right\vert _{one\text{-}column}=\frac{R+1
}{2R}\left( \Delta-R\right) \Delta
\end{equation}
where $\Delta=N_{\psi}$ is the number of boxes in the column with $N_{\psi
}\leq R.$ Indeed this is the correct expression. Such tests show the utility
and power of our general formula.

It takes more effort to compute the general formulas for the
higher Casimir operators in the oscillator approach. However, as
we will see later we only need to compute them for a more
restricted subset of representations within the oscillator
approach, namely those relevant to AdS$_5\times$S$^5$
supergravity. Once we define the algebraic restriction that was
first discovered in the 2T-physics approach, we will easily
determine all Casimir eigenvalues in the last section of this
paper, and thus verify the universal zero.

\section{Lowest weight unitary representations}

The \textquotedblleft ground states\textquotedblright\ that label the lowest
weight representations are singular states which are annihilated by the
bosonic $a\cdot b,$ $\psi\cdot\chi$ and fermionic $a\cdot\chi,$ $\psi\cdot b$
generators. A few examples of such ground states are
\begin{equation}
|0>,\quad\bar{a}_{\alpha}^{n^{\prime}}|0>,\quad\bar{b}^{m^{\prime}\beta
}|0>,\quad\bar{\psi}_{\alpha}^{r^{\prime}}|0>,\quad\bar{\chi}^{s^{\prime
}\alpha}|0>,\quad
\end{equation}%
\begin{equation}
\left( \bar{a}_{\alpha}^{n^{\prime}}\bar{b}^{m^{\prime}\beta}-\frac{1}{C}
\delta_{\alpha}^{\,\beta}\bar{a}^{n^{\prime}}\cdot\bar{b}^{m^{\prime}}
\right) |0>,\quad\bar{a}_{\alpha_{1}}^{n_{1}^{\prime}}\cdots\bar{a}
_{\alpha_{k}}^{n_{k}^{\prime}}|0>,\quad etc.
\end{equation}
In general, the Fock space states that have only upper color indices, or
only lower color indices, are always ground states. When there are both
upper and lower color indices, if they form a traceless tensor in color
space, then the state is also a ground state, provided there are no overall
factors of trace multiplying the expression (i.e. in the examples above $
\bar{a}^{n^{\prime}}\cdot\bar{b}^{m^{\prime}},\bar{a}^{n^{\prime}}\cdot\bar{
\chi}^{s^{\prime}}$ etc. should not appear as factors applied on these
states). Such ground states are classified as irreducible representations of
SU$\left( C\right) \times $U$\left( 1\right) $ and they correspond to Young
tableaux for the color group. The traceless color Young tableau combines a
Young tableau $Y_{a}$ or $Y_{\psi}$ for the color indices on the oscillators
$a$ or $\psi,$ with a Young tableau $Y_{b}^{\ast}$ or $Y_{\chi}^{\ast}$ for
the color indices on the oscillators $b$ or $\chi,$ into an irreducible
traceless color tableau.

The resulting color-traceless tableau determines an eigenvalue for the
quadratic Casimir operator $C_{2}\left( SU\left( C\right) \right) $ and this
number enters in our formulas in Eqs.(\ref{C2},\ref{C2equal}). The other
number $\Delta$ that enters in our formula is determined directly from the
total numbers of oscillators $N_{a},N_{b},N_{\psi},N_{\chi}$ in the
\textquotedblleft ground state\textquotedblright.

In addition, the ground states are classified by the Young tableaux of the
subgroup S(U$\left( N\right) \times$ U$\left( M\right) \times$ U$\left(
R\right) \times$ U$\left( S\right) $), but these are not unrelated to the
color-traceless SU$(C)$ tableaux. Namely, for the bosons $a,b$ when the
color indices are associated with a given tableau, the SU$(N)$ or SU$(M)$
indices must have the same shape tableau, up to a complex conjugation.
Likewise, for the fermions $\psi,\chi$ when the color indices are associated
with a given tableau, the SU$(R)$ or SU$(S)$ indices must correspond to the
reflection of the tableau from its diagonals (interchange rows and columns),
again up to a complex conjugation. In addition, the U$\left( 1\right) $
quantum numbers simply correspond to the numbers of oscillators $%
N_{a},N_{b},N_{\psi},N_{\chi }$ applied on the vacuum. We see that the color
tableaux are directly related to the S(U$\left( N\right) \times$U$\left(
M\right) \times$U$\left( R\right) \times$U$\left( S\right) $) tableaux, and
this explains why the quadratic Casimir operator of SU$\left( N,M|R,S\right)
$ is determined by the color Casimir and $\Delta.$

Applying arbitrary polynomials of all the remaining generators on a given SU$%
\left(N,M|R,S\right)$ ground state produces an infinite tower of states that
have the same Casimir eigenvalues as the ground state (since the SU$%
\left(N,M|R,S\right)$ generators are SU$\left( C\right) \times$U$\left(
1\right)$ singlets). Therefore these towers form infinite dimensional
irreducible representations of SU$\left( N,M|R,S\right).$ All states have
positive norm since they are Fock states created by positive norm
oscillators. Hence these irreducible representations are unitary.

A particular set of ground states that are relevant for classifying the
Kaluza-Klein towers of interest in this paper are those that are
\textquotedblleft color\textquotedblright\ singlet as we will see below. In
this case all the states in the representation are also color singlets since
the SU$\left( N,M|R,S\right) $ generators that produce the infinite tower
are themselves color singlets. In field theoretic applications it is of
interest to organize the towers by listing the minimal set of SU$\left(
2,2\right) $ ground states (both bosonic and fermionic) because their
quantum numbers correspond to the labels on fields which form a
supermultiplet in AdS$_{5}$ space in a field theoretic setting. Therefore
more generally we will discuss the SU$\left( N,M\right) $ ground states that
belong to the same supermultiplet of SU$\left( N,M|R,S\right) .$ The SU$%
\left( N,M\right) $ ground states are those that are annihilated by $
a_{n}\cdot b_{m}$; this subset of states provide the quantum numbers of the
fields that form supermultiplets in field theory.

\section{Group theory for AdS$_{5}\times$S$^{5}$ supergravity}

We use the oscillator approach for SU$\left( 2,2|4\right) $ with generators
constructed from oscillators. Thus, we specialize to $R=4$ and $S=0.$ The U$%
\left( 1\right) $ in SU$\left( 2,2\right) $ is interpreted as the AdS
\textquotedblleft energy\textquotedblright\ operator
\begin{equation}
E=N_{a}+N_{b}+2C.  \label{E}
\end{equation}
It is obtained in the closure of $\left[ a\cdot b,\bar{b}\cdot\bar{a}\right]
,$ and is determined by the total number of bosons ($a$'s or $b$'s), and the
number of \textquotedblleft colors\textquotedblright\ $C.$

The SU$\left( 2,2\right) $ ground states are constructed by applying any
number of creation operators $\bar{a}$ and $\bar{b}$ on the vacuum or other
state made up of only the fermions, provided they are in SU$\left( C\right) $
color-traceless combinations
\begin{equation}
\left[ (\left( \bar{a}\right) ^{n_{a}}\left( \bar{b}\right)
^{n_{b}})_{color\,-traceless}\right] \left( \bar{\psi}\right) ^{n_{\psi}}|0>.
\end{equation}
The numbers of oscillators $n_{a},n_{b},n_{\psi}$ are any integers. The SU$%
\left( 2\right) \times$SU$\left( 2\right) $ indices $m,n$ on the oscillators
(which are not shown) are free to be anything; the color indices $\alpha$
can also be anything as long as they are in traceless combinations when
contracted with $\delta_{\alpha}^{\beta}$ (respecting upper/lower indices).
When the ladder down operators $a_{m}\cdot b_{n}$ are applied on these
states one obtains zero either because there are not enough creation
operators of either kind, or because one is forced to contract the color
indices but the color trace vanishes.

The overall color of the state can be classified by SU$\left( C\right) $
Young tableaux which have definite eigenvalues with respect to the SU$\left(
C\right) $ Casimir operators. The important factor is to include only those
tableau that come from the traceless condition specified withing the $a,b$
oscillators. The total SU$\left( C\right) $ Young tableaux thus obtained
(including the $\bar{\psi}^{\prime}s$) determine the SU$\left( C\right) $
quadratic casimir eigenvalues, and through Eq.(\ref{C2equal})), also the SU$%
\left( 2,2|4\right) $ quadratic casimir eigenvalues.

There are an infinite number of such color-traceless SU$\left( 2,2\right) $
lowest states. Some examples are
\begin{align}
\left( \bar{a}_{n_{1}}^{\alpha_{1}}\cdots\bar{a}_{n_{k}}^{\alpha_{k}}\right)
\left( \bar{\psi}^{r_{1}\beta_{1}}\cdots\bar{\psi}^{r_{l}\beta_{l}}\right)
|0 & >,\quad  \label{lwsu22} \\
\left( \left( \bar{a}\right) _{n}^{\alpha_{1}}\left( \bar{b}\right)
_{m\alpha_{2}}-\frac{1}{C}\delta_{\alpha_{2}}^{\alpha_{1}}\,\bar{a}_{n}\cdot
\bar{b}_{m}\right) \left( \bar{\psi}^{r_{1}\beta_{1}}\cdots\bar{\psi }%
^{r_{l}\beta_{l}}\right) |0 & >,~etc.
\end{align}
The AdS-energy $E$ together with the SU$\left( 2\right) \times$SU$\left(
2\right) \times$SU$\left( 4\right) $ quantum numbers of such lowest states
identify the quantum numbers of the corresponding field in supergravity.
These quantum numbers are easy to figure out by inspecting the state: first
the color indices on the states need to be reduced to irreducible SU$\left(
C\right) $ Young tableau. This forces the SU$\left( 2\right) \times $SU$%
\left( 2\right) \times$SU$\left( 4\right) $ quantum numbers to appear in
definite Young tableaux for each one of these groups. Then, the AdS-energy
is given by the numbers $N_{a}+N_{b}$ of $a^{\prime}s$ and $b^{\prime}s$ as
in Eq.(\ref{E}). The SU$\left( 2\right) \times$SU$\left( 2\right) $ quantum
numbers are also determined as $\left( j_{1},j_{2}\right) $ by counting the
number of unpaired boxes in each of the SU$\left( 2\right) \times$SU$\left(
2\right) $ tableau and dividing by 2. Finally the SU$\left( 4\right) $
quantum numbers are read off from the corresponding Young tableau.

We want to identify the SU$\left( 2,2|4\right) $ lowest states, which when
decomposed into SU$\left( 2,2\right) $ lowest states, contain $\left(
j_{1},j_{2}\right) $ which do not fall outside of the following list%
\begin{align}
\left( j_{1},j_{2}\right) & :\;\left( 0,0\right) ,~\left( 1/2,0\right)
,~\left( 0,1/2\right) ,~\left( 1/2,1/2\right) ,~  \label{j1j2} \\
& \left( 1,0\right) ,~~\left( 0,1\right) ,~\left( 1/2,1\right) ,~\left(
1,1/2\right) ,~\left( 1,1\right) .  \notag
\end{align}
This because the total spin of the field is given by $j_{1}+j_{2},$ and this
should not exceed 2. The representation $\left( j_{1},j_{2}\right) =\left(
1,1\right) $ \ of SU(2)$\times$SU(2) $\subset$ SU$\left( 2,2\right) $
represents the spin 2 graviton with $\left( 2j_{1}+1\right) \times\left(
2j_{2}+1\right) =$3$\times$3=9 independent components, corresponding to the
traceless symmetric field $g_{\mu\nu}$ ($\frac{1}{2}4\times5-1=9$).

The condition of maximum spin 2 comes from the fact that in the
compactification of supergravity higher spinning fields cannot occur.
Therefore we need to identify the SU$\left( 2,2|4\right) $ representations
that have the property of Eq.(\ref{j1j2}). This means that the subgroup SU$%
\left( 2,2\right) $ \textit{lowest state} can have at the most $\left( \bar{%
a }\right) ^{2}\left( \bar{b}\right) ^{2} $ in \textit{color-traceless}
combination applied on the vacuum (with any number of $\bar{\psi}).$
Therefore the lowest state of the full SU$\left( 2,2|4\right) $ should have
the property that it should vanish if more than two powers of the
supergenerators $\bar{\psi}\cdot\bar{a}$ or $\bar{b}\cdot\psi$ are applied,
\textit{provided the color-traceless condition is imposed} on the $a,b$
color indices. We will identify below a unique color singlet SU$\left(
2,2|4\right) $ ground state that satisfies these requirements.

To describe it we examine the lowest SU$\left( 2,2|4\right) $ color singlet
constructed by using the color Levi-Civita symbol. We denote this state
symbolically by $\bar{\psi}^{2C}|0>$%
\begin{equation}
\bar{\psi}^{2C}|0>\sim\left( \varepsilon_{\alpha_{1}\cdots\alpha_{C}}\bar{
\psi}^{a_{1}\alpha_{1}}\cdots\bar{\psi}^{a_{C}\alpha_{C}}\right) \left(
\varepsilon_{\beta_{1}\cdots\beta_{C}}\bar{\psi}^{b_{1}\beta_{1}}\cdots \bar{
\psi}^{b_{C}\beta_{C}}\right) |0>.  \label{bottom}
\end{equation}
After taking into account the fact that the $\bar{\psi}^{a\alpha}$ are
fermions, one deduces that the SU$\left( 4\right) $ indices on the $\bar{%
\psi }^{\prime}s$ must have the permutation properties of the SU$\left(
4\right) $ Young tableau with two rows, each having $C$ boxes. The SU$\left(
4\right) $ Young tableaux labels are therefore $\left( C,C,0,0\right) .$
Since this is a Lorentz singlet, the tower based on this SU$\left(
2,2\right) $ lowest state represents the AdS$_{5}\times$S$^{5}$ scalar field
with appropriately symmetrized/antisymmetrized indices according to the $%
\left( C,C,0,0\right) $ SU$\left( 4\right) $ Young tableau
\begin{equation}
\bar{\psi}^{2C}|0>\rightarrow\phi^{(a_{1}\cdots a_{C}),(b_{1}\cdots
b_{C})}\left( x^{\mu},u\right) \sim\left( C,C,0,0\right) .
\end{equation}
The maximum number of supergenerators that can be applied on this state
before annihilating it are $2C$ for type $\bar{b}\cdot\psi$ supergenerator,
and $2C$ for type $\bar{\psi}\cdot\bar{a}$ supergenerator (since products of
more than $4C$ fermions $\bar{\psi}$ vanish). However when we apply powers
of the supergenerators $\left( \psi\cdot\bar{a}\right) ^{k}\left( \bar{b}
\cdot \bar{\psi}\right) ^{l}$ on the state in Eq.(\ref{bottom}) we create SU$%
\left( 2,2\right)$ states that may have color trace (i.e. contain the color
dot product $\bar{a}\cdot\bar{b}$). These should not be counted among the SU$%
\left( 2,2\right) $ lowest states that identify fields since they are
descendants of the form $\left( \bar{a}\cdot\bar{b}\right)
^{n}|lowest\;SU\left( 2,2\right) >$. Descendants correspond to derivatives
applied on the fields. To identify the fields themselves uniquely one must
project to the color traceless states only.

Thus, the following states obtained by applying supergenerators which form
color traceless combinations
\begin{equation}
\left( \left( \psi\cdot\bar{a}\right) ^{k}\left( \bar{b}\cdot\bar{\psi }%
\right) ^{l}\right) _{color\,traceless}\bar{\psi}^{2C}\,|0>  \label{lowest}
\end{equation}
are also annihilated by $a\cdot b$ thanks to the color-traceless condition.
With this condition we find that the limits of Eq.(\ref{j1j2}) for the spins
of the fields are indeed satisfied because then the powers $k$ or $l$ are
limited to a range $k\leq k_{\max},$ $l\leq l_{\max}$ and $k_{\max}+l_{\max
}\leq4.$ This collection of SU$\left( 2,2\right) $ lowest states - which
result from a single SU$\left( 2,2|4\right) $ lowest state - form
supermultiplets that correspond to the fields on the AdS space $\left(
x^{\mu},u\right) $.

Including all derivatives of the fields corresponds to including all the
states, without the color traceless condition, obtained by applying all the
generators with arbitrary powers. In our construction applying all powers of
generators without restrictions corresponds to a unitary representation
based on the lowest SU$\left( 2,2|4\right) $ state. By construction, all
states are automatically color singlets for any number of colors $C,$ since
in Eq.(\ref{lowest}) $\bar{\psi}^{2C}$ as well as all SU$\left( 2,2|4\right)
$ generators are color singlets.\ The towers based on the lowest
supermultiplets in Eq.(\ref{lowest}) correspond to the fields that form
supermultiplets without derivatives on the fields$.$ The quantum numbers on
the fields are the SU$\left( 2\right) \times$SU$\left( 2\right) \times $SU$
\left( 4\right) $ Young tableau labels of the lowest states. By
construction, these fields form irreducible supermultiplets of SU$\left(
2,2|4\right) .$ There is such a supermultiplet for every value of the color $%
C=1,2,3,\cdots\infty$ with its lowest state being $\bar{\psi}^{2C}\,|0>.$

One important fact that should be noted is that the color Casimir eigenvalue
$C_{2}^{SU\left( C\right) }$ of these color singlet states is automatically
zero!

\subsection{Yang-Mills supermultiplet}

We now discuss SU$\left( 2,2|4\right) $ with only one color $C=1.$ This case
is simple because there are no color traceless combinations to discuss.
Consider the lowest state $\bar{\psi}^{2}|0>.$ More than two powers of the
supergenerators $\bar{b}\bar{\psi}$ annihilate this state since there are
only four different $\bar{\psi}^{a}.$ Also more than two powers of $\psi\bar{%
a}$ annihilate this state since no more than two annihilation operators $%
\psi_{a}$ can survive on this state. All the distinct SU$\left( 2,2\right) $
lowest states are obtained by applying all powers of the supergenerators
\begin{align}
\bar{\psi}^{2}|0 & >,\quad\phi^{\lbrack a_{1}a_{2}]},\quad\left(
scalar,6\right)  \label{c1scalar} \\
\left( \psi\bar{a}\right) \bar{\psi}^{2}|0 &
>,\quad\varepsilon^{a_{1}a_{2}a_{3}a_{4}}\,\chi_{a_{4}},\quad\left(
left\,spinor,4\right) \\
\left( \psi\bar{a}\right) ^{2}\bar{\psi}^{2}|0 & >,\quad F_{\left( \mu
\nu\right) _{+}}\,\varepsilon^{a_{1}a_{2}a_{3}a_{4}},\quad\left(
self\,dual\,tensor,1\right) \\
\left( \bar{b}\bar{\psi}\right) \bar{\psi}^{2}|0 &
>,\quad\xi^{a},\quad\left( right\,spinor,4^{\ast}\right) \\
\left( \bar{b}\bar{\psi}\right) ^{2}\bar{\psi}^{2}|0 & >,\quad F_{\left(
\mu\nu\right) _{-}},\quad\left( anti\,\,self\,dual\,tensor,1\right)
\end{align}
These are all annihilated by $\left( ab\right) $ since they are of the type (%
\ref{lwsu22}). The irreducible SU$\left( 2,2\right) $ modules (infinite
towers of descendants) are obtained by applying all powers of $\left( \bar
{%
b}\bar{a}\right) $ on each of these lowest states. There are no other SU$%
\left( 2,2\right) $ lowest states in the Hilbert space based on $\bar{\psi}%
^{2}|0>$ because states obtained by applying higher powers of
supergenerators, such as $\left( \psi\bar{a}\right) \left( \bar{b}\bar {\psi}%
\right) \bar{\psi}^{2}|0>$ can be brought to the form $\left( \bar
{b}\bar{a%
}\right) \bar{\psi}^{2}|0>$ and therefore already are accounted in the
towers including descendants based on $\bar{\psi}^{2}|0>.$

Thus, the lowest state $\bar{\psi}^{2}|0>$ fully characterizes the entire SU$%
\left( 2,2|4\right) $ supermultiplet. In more detail, this state has the
form $\bar{\psi}^{a_{1}}\bar{\psi}^{a_{2}}|0>$ and therefore is the 6
dimensional antisymmetric tensor of SU$\left( 4\right) ,$ and is a scalar
under SU$\left( 2,2\right) ,$ as indicated in Eq.(\ref{c1scalar}). Therefore
the SU$\left( 2,2\right)$ tower based on this state represents the scalar
fields $\phi^{\lbrack a_{1}a_{2}]}\left( x^{\mu},u\right) $ on AdS space.
The tower corresponds to the derivatives of the field that form a complete
basis$.$ Similarly, in terms of the Lorentz group embedded in SU$\left(
2,2\right) $ the other lowest states give spin 1/2 and spin 1 fields as
indicated in the equations above. This collection of states, which make an
irreducible multiplet of SU$\left( 2,2|4\right) $ may be viewed as the
scalar supermultiplet of $N=4$ supersymmetry. This supermultiplet does not
contain a spin 2 field since the maximum spin among the lowest states is 1.
This set of fields correspond to the super Yang-Mills theory.

\section{SU$\left( 2,2|4\right) $ KK towers in supergravity}

For two colors, $C=2,$ the Lorentz and SU$\left( 4\right) $ content of the
fields are as follows. The Lorentz part is decribed by $\left[ j_{1},j_{2}%
\right] $ that label the SU$\left( 2\right) \times$SU$\left( 2\right)
\subset $SU$\left( 2,2\right) .$ By construction of the state, $\left[
j_{1},j_{2}\right] $ are obtained by using angular momentum addition rules
that combine the 1/2 spins carried by the labels $\bar{a}$,$\bar{b}.$ The SU$%
\left( 4\right) $ content is given by U$\left( 4\right) $ Young tableaux
with 4 rows with $\left( n_{1},n_{2},n_{3},n_{4}\right) $ boxes
respectively. The $n_{i}$ are obtained by combining the Young tableaux of
the initial state $\bar{\psi}^{2C}|0>,$ which is $\left( C,C,0,0\right) $
with those of the supergenerators after the color traceless condition on the
$a,b$ part. From the Young tableaux one may extract SU$\left( 4\right) $
Dynkin labels consisting of three integers $\left\{
\nu_{1},\nu_{2},\nu_{3}\right\} \ $ given by $\nu_{1}=n_{1}-n_{2},$ $%
\nu_{2}=n_{2}-n_{3}$ and $\nu_{3}=n_{3}-n_{4}.$ In what follows we provide
the $\left[ j_{1},j_{2}\right] _{E}$, and SU$\left( 4\right) $ Young and
Dynkin labels, where AdS energy $E=\left( k+l\right) +2C$ appears as a
subscript. We will also give a field theory notation with upper and lower U$%
\left( 4\right) $ indices that correspond to traceless tensors, and Lorentz
indices including vector and dotted and undotted spinor indices\footnote{%
The migration from the $\left( j_{1},j_{2}\right) $ labels for compact SU$%
\left( 2\right) \times $SU$\left( 2\right) \subset$SU$\left( 2,2\right) $ to
the Lorentz labels for noncompact SO$\left( 3,1\right) \subset$SU$\left(
2,2\right) $ corresponds to a change of basis within the same group SU$%
\left( 2,2\right) $ to describe the irreducible representation (the entire
tower)$.$ In terms of oscillators this is obtained by a non-unitary
Bogoliubov transformation \cite{G}.}.

The $C=2$ field content obtained with this approach coincides with the
graviton supermultiplet in AdS$_{5}\times$S$^{5}$ supergravity written in
terms of $d=4$ Lorentz and spinor labels and with $SU(4)$ representations.
The SU$\left( 2,2\right) $ lowest states obtained by applying the
supergenerators on the SU$\left( 2,2|4\right) $ lowest state $\bar{\psi}%
^{2C}|0>$ as in Eq.(\ref{lowest}), satisfy $k+l\leq4$ and the conditions on
spin in Eq.(\ref{j1j2}), for $C=2,$ as follows
\begin{align}
k & =0,\quad l=0:\quad\left[ 0,0\right] _{4}\left( 2,2,0,0\right) \left\{
0,2,0\right\} ,\quad\phi_{\left[ ab\right] }^{\left[ cd\right] } \\
k & =1,\quad l=0:\quad\left[ \frac{1}{2},0\right] _{5}\left( 2,2,1,0\right)
\left\{ 0,1,1\right\} ,\quad\left( \zeta_{\alpha}\right) _{\left[ ab\right]
}^{c} \\
k & =0,\quad l=1:\quad\left[ 0,\frac{1}{2}\right] _{5}\left( 2,1,0,0\right)
\left\{ 1,1,0\right\} ,\quad\left( \bar{\zeta}_{\dot {\alpha}}\right) _{a}^{%
\left[ cd\right] } \\
k & =2,\quad l=0:\left\{
\begin{array}{c}
\left[ 1,0\right] _{6}\left( 2,2,1,1\right) \left\{ 0,1,0\right\} ,\quad
B_{\left( \mu\nu\right) _{+}}^{\left[ ab\right] } \\
\left[ 0,0\right] _{6}\left( 2,2,2,0\right) \left\{ 0,0,2\right\}
,\quad\varphi_{\left( ab\right) }%
\end{array}
\right. \\
k & =0,\quad l=2:\left\{
\begin{array}{c}
\left[ 0,1\right] _{6}\left( 1,1,0,0\right) \left\{ 0,1,0\right\} ,\quad
B_{\left( \mu\nu\right) _{-}}^{\left[ ab\right] } \\
\left[ 0,0\right] _{6}\left( 2,0,0,0\right) \left\{ 2,0,0\right\} ,\quad\bar{%
\varphi}^{\left( ab\right) }%
\end{array}
\right. \\
k & =1,\quad l=1:\quad\left[ \frac{1}{2},\frac{1}{2}\right] _{6}\left(
2,1,1,0\right) \left\{ 1,0,1\right\} ,\quad\left( A_{\mu}\right) _{b}^{a} \\
k & =3,\quad l=0:\quad\left[ \frac{1}{2},0\right] _{7}\left( 2,2,2,1\right)
\left\{ 0,0,1\right\} ,\quad\left( \xi_{\alpha}\right) _{a} \\
k & =0,\quad l=3:\quad\left[ 0,\frac{1}{2}\right] _{7}\left( 1,0,0,0\right)
\left\{ 1,0,0\right\} ,\quad\left( \bar{\xi}_{\dot{\alpha}}\right) ^{a} \\
k & =1,\quad l=2:\quad\left[ \frac{1}{2},1\right] _{7}\left( 1,1,1,0\right)
\left\{ 0,0,1\right\} ,\quad\left( \Psi_{\mu\dot{\alpha}}\right) _{a} \\
k & =2,\quad l=1:\quad\left[ 1,\frac{1}{2}\right] _{7}\left( 2,1,1,1\right)
\left\{ 1,0,0\right\} ,\quad\left( \bar{\Psi}_{\mu\alpha }\right) ^{a} \\
k & =4,\quad l=0:\quad\left[ 0,0\right] _{8}\left( 2,2,2,2\right) \left\{
0,0,0\right\} ,\quad\Sigma \\
k & =0,\quad l=4:\quad\left[ 0,0\right] _{8}\left( 0,0,0,0\right) \left\{
0,0,0\right\} ,\quad\bar{\Sigma} \\
k & =2,\quad l=2:\quad\left[ 1,1\right] _{8}\left( 1,1,1,1\right) \left\{
0,0,0\right\} ,\quad g_{\mu\nu} \\
k & =3,\quad l=1:\quad excluded~by~color~traceless~rule \\
k & =1,\quad l=3:\quad excluded~by~color~traceless~rule
\end{align}
Of course, this list of supergravity states is in agreement with previous
results of G\"{u}naydin and Marcus in \cite{barsgunaydin} that used
oscillator methods with a different SU$\left( 4\right) $ classification of
oscillators than ours. Although equivalent, in our approach the entire
multiplet is more clearly identified by simply specifying only the $k=l=0$
state $\bar{\psi}^{4}|0>$.

In the same manner the supermultiplets in higher Kaluza-Klein towers for all
values of color $C=2,3,4,\cdots$ are identified with only the lowest state $%
\bar{\psi}^{2C}|0>,$ and the resulting supermultiplets that follow from them
are in agreement with those found by G\"{u}naydin and Marcus, and those of
Kim, Romans and Van Nieuwenhuizen \cite{KRN} who used a completely different
technique.

The significant point in our analysis is that all the states have been
identified as color singlets. This fixes $C_{2}^{SU\left(C\right)}=0,$ while
the value of $\Delta$ is read off immediately from the lowest state $\bar{%
\psi}^{2C}|0>$ as being $\Delta=N_{\psi}=2C$. The value of $\Delta$ is fixed
for the entire tower since the SU$\left( 2,2|4\right)$ generators commute
with $\Delta$. This allows us to compute the SU$\left(2,2|4\right)$ Casimir
eigenvalue for all the states, which is the new point in our investigation.

\subsection{AdS-CFT}

It must be noted that one can apply these ideas in the context of the
AdS-CFT correspondence. The same set of supermultiplets discussed above can
be built by starting from the scalar fields $\phi^{\left[ ab\right] }$ in
N=4 super Yang-Mills theory in four dimensions, as follows.

The scalars $\phi^{\left[ ab\right] }$ are in the 2-index antisymmetric
representation of SU$\left( 4\right) $ which forms the vector of SO$\left(
6\right) ,$ while being in the adjoint of Yang-Mills gauged SU$\left(
N\right) .$ The spacetime and SU$\left( 4\right) $ quantum numbers of this
CFT state should be compared to the $C=1$ lowest state $\bar{\psi}^{2}|0>$
of the Yang-Mills supermultiplet given in Eq.(\ref{c1scalar}).

Using the AdS-CFT correspondence, the AdS$_{5}\times$S$^{5}$ supermultiplets
that we discussed above would emerge by constructing the lowest states in
the super Yang-Mills context. These are obtained by taking the gauge
invariant SU$\left( N\right) $ trace over $C$ such scalars Tr$\left( \phi^{%
\left[ a_{1}b_{1}\right] }\ldots\phi^{\left[ a_{C}b_{C}\right] }\right) ,$
and further demanding that their SU$\left( 4\right) $ quantum numbers be
projected to the Young tableau $\left( C,C,0,0\right).$ A priori there are
gauge singlet combinations of the scalars that do not correspond to the SU$%
(4)$ tableau $\left( C,C,0,0\right)$. Our AdS-CFT correspondence identifies
only this tableau. Then supersymmetry would correctly generate the remainder
of the Kaluza-Klein tower as we discussed above in our oscillator formalism.
In identifying the fields in a supermultiplet by applying super conformal
symmetry, higher derivatives that result from applying the conformal
supergenerators must be dropped (this is the analog of the color traceless
condition in Eq.(\ref{lowest}).

Note that the first power, $C=1,$ is not included among the AdS$_{5}\times$S$%
^{5}$ fields since the gauge invariant SU$\left( N\right) $ trace is zero
for $C=1$, i.e. Tr$\left( \phi^{\left[ a_{1}b_{1}\right] }\right) =0.$

\section{The zero}

The quadratic Casimir eigenvalues for any color singlet states of SU$\left(
2,2|4\right) $ are given in Eq.(\ref{colorless}). Note that generally these
are not zero. However, in our case, since the ground state satisfies $%
\Delta=N_{\psi}=2C,$ we see that the quadratic Casimir vanishes for the
ground state $\bar{\psi}^{2C}|0>$ as well as for the entire Kaluza-Klein
tower for each $C.$ Therefore all the states of AdS$_{5}\times$S$^{5}$
supergravity, or those obtained through the AdS-CFT correspondence with the
prescription given in the previous paragraph, have zero SU$\left(
2,2|4\right) $ quadratic Casimir eigenvalues.

There is no explanation of this fact within supergravity or AdS-CFT. However
this mysterious zero is easily explained in \cite{AdS5S5} as a consequence
of the symmetry structures revealed in 2T-physics. It follows from a
relationship between the fundamental 2T-physics gauge symmetry Sp$\left(
2\right) $ and local kappa symmetry, and the corresponding gauge invariance
condition for physical states. The Sp$\left( 2\right) $ gauge symmetry is
the universal mechanism for the projection of $d+2$ dimensional 2T-physics
theory to a $d$ dimensional 1T-physics theory. The zero SU$\left(
2,2|4\right) $ Casimir follows directly from the gauge invariance of
physical states under this Sp$\left( 2\right) $ symmetry, that is
\begin{equation}
X^{2}=P^{2}=X\cdot P=0
\end{equation}%
where $X^{M},P^{M}$ describe 12-dimensional phase space (see \cite{survey2T}
), and $X^{2},P^{2},X\cdot P$ are the Sp$\left( 2\right) $ generators.

In fact, the vanishing of the quadratic Casimir is a reflection of
a more general universal zero that applies to all the Casimir
eigenvalues for all the Kakuza-Klein towers in AdS$_{5}\times
$S$^{5}$ supergravity. As shown in \cite{AdS5S5}, the generators
of SU$\left( 2,2|4\right) $ that describe these states can be
written in the form of an 8$\times 8$ supertraceless matrix $J$
constructed from the 12-dimensional super phase space $\left(
X^{M},P^{M},\Theta \right) $ in a 2T formulation of a
superparticle in 12-dimensions, with 32 fermions $\Theta $'s. This
provides a dynamical particle representation of the $J$ in
Eq.(\ref{J}) after modifying it for $M+N=R+S$ (this means
supertraceless $J$ for oscillators is defined by imposing
Str$J=\Delta+(N-R)C=0$). The particle representation of the
supermatrix has the algebraic property \cite{AdS5S5}
\begin{equation}
\left( J\right) ^{2}=\frac{1}{4}\hbar ^{2}l\left( l+4\right)
\mathbf{1-} 2\hbar \left( J\right) ,  \label{JJ}
\end{equation}%
where the first term is proportional to the identity supermatrix
$\mathbf{1,} $ with $l=0,1,2,\cdots $ , where $l$ labels the
harmonics on $S^{5}\mathbf{.} $ The representation of SU$\left(
2,2|4\right) $ changes as $l$ changes, corresponding to the
supergravity Kaluza-Klein towers. We have verified that the
oscillator representations described in this paper, in the color
singlet sector, satisfy precisely these constraint equations
predicted by the 2T-physics 12-dimensional structure. The details
will be further elaborated in a future publication.

Thus Eq.(\ref{JJ}) is the fundamental algebraic structure that
defines the supergravity Kaluza-Klein fields. It selects the
correct subset of physical states in the vast Fock space of
oscillators. We conjecture that the SU$\left( 2,2|4\right)$
representations that satisfy these conditions for the generators
$J$ are unique in any formalism. These algebraic equations are
understood naturally as the physical state constraint equations
that follow from local supersymmetries for the 12D superparticle
described in \cite{AdS5S5}.

Higher powers of the supermatrix $\left( J\right) ^{n}$ can now be
computed by repeated applications of this formula. In particular
the quadratic and all higher Casimir operators of SU$\left(
2,2|4\right) $ must vanish in these realizations
since the supertrace of $\mathbf{1}$\textbf{\ }and the supertrace of\textbf{%
\ \ }$J$\textbf{\ }are zero
\begin{equation}
C_{n}\left( 2,2|4\right) =Str\left( J^{n}\right) =0.  \label{Cn}
\end{equation}%
These SU$\left( 2,2|4\right) $ properties arose through constraints
(associated with gauge symmetries) on a 12-dimensional super phase space,
and hence the properties of these representations reflect the underlying
12-dimensional structure.

In this paper we have seen that the universal zero Casimirs
prediction implied by 2T-physics \cite{AdS5S5} is a group
theoretical fact in supergravity or the AdS-CFT correspondence, by
using traditional group theoretical methods involving the
oscillator representations.

It must be mentioned that the 2T-physics superparticle approach
does not predict a similar universal zero for standard
AdS$_4\times$S$^7$ or AdS$_7\times$S$^4$ supergravities, therefore
this should not be expected. However, it does make certain
predictions \cite{AdS5S5} for certain structures related to such
spaces which again can in principle be verified.

The 2T-physics approach provides a much simpler view of the SU$\left(
2,2|4\right) $ and Sp$\left( 2\right) $ covariant dynamics, and their
inter-relationship. The properties of SU$\left( 2,2|4\right) $
representations discussed in this paper followed from the quantum states of
a superparticle, therefore they apply to supergravity, but not to higher
superstring states. The extension of the 2T-physics superparticle
formulation to superstrings would be very interesting, and would be expected
to reveal profound structures.


\begin{thebibliography}{9}
\bibitem{super2t} I. Bars, C. Deliduman and D. Minic, \textquotedblleft
Supersymmetric Two-Time Physics\textit{\textquotedblright, }Phys. Rev.
\textbf{D59} (1999) 125004, hep-th/9812161; \textquotedblleft Lifting
M-theory to Two-Time Physics\textit{\textquotedblright, }Phys. Lett. \textbf{%
\ B457} (1999) 275, hep-th/9904063; \newline
I. Bars, \textquotedblleft\textit{2}T formulation of superconformal dynamics
relating to twistors and supertwistors\textquotedblright, Phys. Lett.
\textbf{B483 }(2000) 248, hep-th/0004090; \textquotedblleft A Toy
M-model\textquotedblright, in preparation (partial results in hep-th/9904063
and hep-th/0008164).

\bibitem{survey2T} I. Bars, reviews in conference proceedings:
\textquotedblleft Two-Time Physics\textquotedblright, hep-th/9809034;
\textit{\textquotedblleft}Survey of Two-Time Physics\textit{\
\textquotedblright} , hep-th/0008164; \textquotedblleft2T-Physics
2001\textquotedblright, hep-th/0106021.

\bibitem{AdS5S5} I. Bars, \textquotedblleft Hidden 12-dimensional Structures
in AdS$_{5}\times$S$^{5}$ and M$^4\times$R$^6$ Supergravities",
hep-th/0208012.

\bibitem{barsgunaydin} I. Bars and M. G\"{u}naydin, Comm. Math. Phys.
\textbf{91 }(1983) 31. M. G\"{u}naydin and N. Marcus, Class. and Quant.
Grav. \textbf{2 }(1985) L11; M. G\"{u}naydin, D. Minic and M. Zagermann,
Nucl. Phys. \textbf{B534 }(1998) 96; Nucl. Phys. \textbf{B544 }(1999) 737.
M. G\"{u}naydin and D. Minic, Nucl. Phys. \textbf{B523 }(1998) 145.

\bibitem{BB} I. Bars and A.B. Balantekin, J. Math. Phys. \textbf{22} (1981)
1149; J. Math. Phys. \textbf{22} (1981)1810.

\bibitem{GS} M. G\"{u}nayd\i n and C. Sa\c{c}l\i o\~{g}lu, Comm. Math. Phys.
\textbf{87} (1982) 159.

\bibitem{G} M. G\"{u}nayd\i n, \textquotedblleft AdS/CFT dualities and
\ldots\ Wigner versus Dirac\textquotedblright, hep-th/0005168.

\bibitem{KRN} H.J.\ Kim, L.J. Romans and P. van Nieuwenhuizen, Nucl. Phys.
\textbf{B242} (1984) 377.
\end{thebibliography}
\end{document}